\documentclass[a4paper]{jpconf}
\usepackage{amsfonts}
\usepackage{graphicx}
\usepackage{amssymb,latexsym}

\def\be{\begin{equation}}
\def\ee{\end{equation}}
\def\bq{\begin{eqnarray}}
\def\eq{\end{eqnarray}}
\def\beq{\begin{eqnarray*}}
\def\eeq{\end{eqnarray*}}

\def\r{\rho}
\def\t{\tau}

\def\bfp{\mathbf{p}}
\def\bfA{\mathbf{a}}

\begin{document}

\title{The Dominant Balance at Cosmological Singularities}
\author{Spiros Cotsakis$^{1,\dagger}$ and John D. Barrow$^{2,\ddagger}$}

\address{$^1$Research Group of Mathematical Physics and Cosmology,
Department of Information and Communication Systems Engineering,
University of the Aegean, Karlovassi 83 200, Samos, Greece} 
\address{$^2$ DAMTP, Centre for Mathematical
Sciences, University of Cambridge,
Wilberforce Road,
Cambridge CB30WA, UK} \ead{skot@aegean.gr$\dagger$} %
\ead{J.D.Barrow@damtp.cam.ac.uk$\ddagger$}

\begin{abstract}
\noindent We define the notion of a finite-time singularity of a vector
field and then discuss a technique suitable for the asymptotic analysis of
vector fields and their integral curves in the neighborhood of such a
singularity. Having in mind the application of this method to cosmology, we
also provide an analysis of the time singularities of an isotropic universe
filled with a perfect fluid in general relativity.
\end{abstract}
\section{Introduction}

There are two approaches to characterizing spacetime singularities in a
cosmological context. The first approach may be called \emph{geometric} and
consists of finding sufficient and/or necessary conditions for singularity
formation, or absence, \emph{independently} of any specific solution of the
field equations under general conditions on the matter fields. Methods of
this sort include those based on an analysis of geodesic congruences in
spacetime and lead to the well known singularity theorems, cf. \cite{he73},
as well as those which are depend on an analysis of the geodesic equations
themselves and lead to completeness theorems such as those expounded in \cite%
{cbc02}, and the classification of singularities in \cite{ck06}.  

The second approach to the singularity problem can be termed \emph{dynamical}
and refers to characterizing cosmological singularities in a geometric
theory of gravity by analysing the dynamical field equations of the theory%
\emph{\ } It uses methods from the theory of dynamical systems and can be 
\emph{global}, referring to the asymptotic behaviour of the system of field
equations for large times, or \emph{local}, giving the behaviour of the
field components in a small neighborhood of the finite-time singularity.

In this latter spirit, we present here a local method for the
characterization of the asymptotic properties of solutions to the field
equations of a given theory of gravity in the neighborhood of the spacetime
singularity\footnote{%
Similar approaches have been used in the past to test given systems of
equations for integrability under the so-called Painleve test. However, our
approach is more geometric in nature and is not related to integrability.}.
We are interested in providing an asymptotic form for the solution near 
singularities of the gravitational field and understanding all possible
dominant features of the field as we approach the singularity. We call this
approach the \emph{method of asymptotic splittings.}

In the following sections, we give an outline of the method of asymptotic
splittings with a view to its eventual application to cosmological
spacetimes in different theories of gravity. For the sake of illustration,
in the last Section we analyze the asymptotic behaviour of a
Friedmann-Robertson-Walker (FRW) universe filled with perfect fluid in
Einstein's general relativity, which provides the simplest, nontrivial
cosmological system.

\section{Finite-time singularities: Definitions}

It is advantageous to work on any differentiable manifold $\mathcal{M}^{n}$,
although for specific applications we restrict attention to open subsets or $%
\mathbb{R}^{n}$ We shall use interchangeably the terms vector field $\mathbf{%
f}:\mathcal{M}^{n}\rightarrow \mathcal{TM}^{n}$ and dynamical system defined
by $\mathbf{f}$ on $\mathcal{M}^{n}$, $\dot{\mathbf{x}}=\mathbf{f}(\mathbf{x}%
)$, with $(\cdot )\equiv d/dt$. Also, we will use the terms 'integral curve' 
$\mathbf{x}(t,\mathbf{x}_{0})$ of the vector field $\mathbf{f}$ with initial
condition $\mathbf{x}(0)=\mathbf{x}_{0}$, and 'solution' of the associated
dynamical system $\mathbf{x}(t;\mathbf{x}_{0})$ passing through the point $%
\mathbf{x}_{0}$, with identical meanings.

Given a vector field $\mathbf{f}$ on the $n$-dimensional manifold $\mathcal{M%
}^{n}$, we define the notion of a \emph{general} solution of the associated
dynamical system as a solution that depends on $n$ arbitrary constants of
integration, $\mathbf{x}(t;\mathbf{C}),\,\mathbf{C}=(c_{1},\cdots ,c_{n})$.
These constants are uniquely determined by the initial conditions in the
sense that to each $\mathbf{x}_{0}$ we can always find a $\mathbf{C}%
_{0}=(c_{10},\cdots ,c_{n0})$ such that the solution $\mathbf{x}(t;\mathbf{C}%
_{0})$ is the unique solution passing through the point $\mathbf{x}_{0}$.
Therefore, a property holds \emph{independently} of the initial conditions
if and only if it is a property of a general solution of the system.

A \emph{particular} solution of the dynamical system is any solution
obtained from the general solution by assigning specific values to at least
one of the arbitrary constants. The particular solutions containing $k$
arbitrary constants can be viewed as describing the evolution in time of
sets of initial conditions of dimension $k$ strictly smaller than $n$. A
particular solution is called an \emph{exact} solution of the dynamical
system when $k=0$\footnote{%
There are solutions, called \emph{singular} solutions, which have the
property that a certain jacobian vanishes on them, and so they are not
obtainable from the general solution of the system like the particular
solutions. Below we shall use this term but with a totally different meaning
as our notion of a solution with a time singularity (which can be also
called a singular solution) is completely different.}. Thus, in our
terminology, a particular solution is a more general object than any exact
solution, the latter having the property that all arbitrary constants have
been given specific values. The hierarchy: exact (no arbitrary constants) to
particular (strictly less than maximum number of arbitrary constants) to
general solutions, will play an important role in what follows.

General, particular, or exact solutions of dynamical systems can develop 
\emph{finite-time singularities}; that is, instances where a solution $%
\mathbf{x}(t;c_{1},\cdots ,c_{k}),\,k\leq n$, misbehaves at a finite value $%
t_{\ast }$ of the time $t$. This is made precise as follows. We say that the
system $\dot{\mathbf{x}}=\mathbf{f}(\mathbf{x})$ (equivalently, the vector
field $\mathbf{f}$) has a \emph{finite-time singularity} if there exists a $%
t_{\ast }\in \mathbb{R}$ and a $\mathbf{x}_{0}\in \mathcal{M}^{n}$ such that
for all $M\in \mathbb{R}$ there exists an $\delta >0$ such that 
\begin{equation}
||\mathbf{x}(t;\mathbf{x}_{0})||_{L^{p}}>M,  \label{sing1}
\end{equation}%
for $|t-t_{\ast }|<\delta $. Here $\mathbf{x}:(0,b)\rightarrow \mathcal{M}%
^{n}$, $\mathbf{x}_{0}=\mathbf{x}(t_{0})$ for some $t_{0}\in (0,b)$, and $%
||\cdot ||_{L^{p}}$ is any ${L^{p}}$ norm. We say that the vector field has
a \emph{future} (resp. \emph{past}) singularity if $t_{\ast }>t_{0}$ (resp. $%
t_{\ast }<t_{0}$). Note also, that $t_{0}$ is an arbitrary point in the
domain $(0,b)$ and may be taken to mean `now'. Alternatively, we may set $%
\tau =t-t_{\ast }$, $\tau \in (-t_{\ast },b-t_{\ast })$, and consider the
solution in terms of the new time variable $\tau $, $\mathbf{x}(\tau ;%
\mathbf{x}_{0})$, with a finite-time singularity at $\tau =0$. We see that
for a vector field to have a finite-time singularity there must be at least
one integral curve passing through the point $\mathbf{x}_{0}$ of $\mathcal{M}%
^{n}$ such that at least one of its ${L^{p}}$ norms diverges at $t=t_{\ast }$%
. We write 
\begin{equation}
\lim_{t\rightarrow t_{\ast }}||\mathbf{x}(t;\mathbf{x}_{0})||_{L^{p}}=\infty
,  \label{sing2}
\end{equation}%
to denote a finite-time singularity at $t_{\ast }$.

One of the most interesting problems in the theory of singularities of
vector fields is to find the structure of the set of points $\mathbf{x}_{0}$
in $\mathcal{M}^n$ such that, when evolved through the dynamical system
defined by the vector field, the integral curve of $\mathbf{f}$ passing
through a point in that set satisfies property (\ref{sing2}).

Another important question, of special interest in relativistic cosmology,
is to discover the precise relation between the finite-time singularities of
vector fields that arise as reductions of the field equations and those that
emerge in the form of geodesic incompleteness. The difficulty here is that
the finite-time singularities of vector fields appear to be unconnected to
geodesic incompleteness and conversely, singularities which arise through
the formation of conjugate points do not seem to demand, or require, any
dynamical description.

\section{Fixed and movable singularities}

Finite-time singularities of (general or particular) solutions of linear
dynamical systems are located at the singularities of their coefficients and
are \emph{fixed} because they are known from the singularities of the
coefficients of the system. The fixed singularities in a solution are
therefore independent of the choice of initial conditions. In contrast,
solutions of nonlinear systems can develop finite-time singularities that
are either fixed or movable. A singularity is \emph{fixed} if it is a
singularity of $\mathbf{x}(t;\mathbf{C})$ for all $\mathbf{C}$; otherwise,
we say it is a \emph{movable singularity}.

Note that any fixed finite-time singularity of a particular solution ($k<n$)
cannot be a fixed singularity of a general solution since at least one of
the constants appearing in the general solution has been set to zero, and so
this singularity is not one of a $\mathbf{x}(t;\mathbf{C})$ for \emph{all} $%
\mathbf{C}$, that is independent of the initial conditions. Hence, fixed
finite-time singularities \emph{in a general solution} cannot be understood
by studying fixed singularities in particular solutions. However, movable
singularities of a particular solution, if they \emph{are} singularities (in
the sense of the definition above) of the general solution, will always be
movable ones. Therefore, movable finite-time singularities in particular
solutions make a nonzero contribution to the singularity pattern of the
vector field and must be taken into consideration in the general study of
its singularities.

It may also happen that a dynamical system has no movable singularities in
the general solution but still has a singular or particular solution with a
movable singularity. Hence, a fixed (or movable) singularity in a general
solution can be a fixed (or movable) singularity of some particular solution
but not vice versa. In general,  movable singularities are more interesting,
since the issue of choosing initial conditions plays an important role for
them; consequently, we shall restrict our attention to them almost
exclusively in what follows.

How should we tackle the geometric problem of describing the behaviour of
vector fields and their integral curves -- solutions of the associated
dynamical system -- in the neighborhood of a finite-time movable
singularity? Assume that we are given a vector field, and we know that at
some point, $t_{\ast }$, a system of integral curves, corresponding to a
particular or a general solution, has a (future or past) finite-time
singularity in the sense of definition (\ref{sing1}). The approach we take
in this paper is an asymptotic one. The vector field (or its integral
curves) can basically do two things sufficiently close to the finite-time
singularity, namely, it can either show some dominant feature or not. In the
latter case, the integral curves can `spiral' in some way around the
singularity \emph{ad infinitum} so that (\ref{sing1}) is satisfied and the
dynamics are totally controlled by the subdominant (lower-order in terms of
weight - see below) terms, whereas in the former case solutions share a
distinctly dominant behaviour on approach to the singularity at $t_{\ast }$
determined by the most nonlinear terms.

To describe both cases invariably, we decompose the vector field into
simpler, component vector fields and examine whether the most nonlinear one
of these shows a dominant behaviour while the rest become subdominant in
some exact sense. Using this picture, we then built a system of integral
curves corresponding, where feasible, to the general solution, and sharing
exactly its characteristics in a sufficiently small neighborhood of the
finite-time singularity. The construction of these solutions is given as a
formal asymptotic series expansion around the singularity and is done
term-by-term.

\section{Weight-homogeneous vector fields}

We begin by introducing some useful notation and terminology. We write 
\begin{equation}  \label{scale function}
\mathbf{x}(\tau) =\mathbf{a}\tau^{\mathbf{p}},
\end{equation}
to denote the function 
\[
\mathbf{x}(\tau) =\left(\alpha\tau^p ,\beta\tau^q,\gamma\tau^r,\cdots
\right), 
\]
where $\mathbf{a}=(\alpha,\beta,\gamma,\cdots)\in\mathbb{C}^n\setminus\{%
\mathbf{0}\},$ and $\mathbf{p}=(p,q,r,\cdots)\in\mathbb{Q}^n.$ A function of
the form (\ref{scale function}) is \emph{scale invariant} in the sense that
a change in the time scale, $\tau^{\prime}=\eta\tau$, $\eta\in\mathbb{R}$,
reveals that we must also have $\mathbf{x}^{\prime}=\eta^{\mathbf{p}}\mathbf{%
x}$, and conversely.

Now we demand that a scale invariant function $\mathbf{x}(\tau ;\mathbf{x}%
_{0})$ of the form (\ref{scale function}) is an integral curve of the vector
field $\mathbf{f}$ passing through $\mathbf{x}_{0}$, or equivalently, the
associated dynamical system $\dot{\mathbf{x}}=\mathbf{f}(\mathbf{x})$ has a
particular solution that is scale invariant, valid in a neighborhood of the
assumed finite-time singularity. This means that 
\begin{equation}
\mathbf{a}\mathbf{p}\tau ^{\mathbf{p}-\mathbf{1}}=\mathbf{f}(\mathbf{a}\tau
^{\mathbf{p}}).  \label{basic1}
\end{equation}%
The notation $\mathbf{a}\mathbf{p}$ stands for the monomial $a_{i}p_{i}$ and
is valid for each $i=1,\cdots ,n$ (range convention, no summation).

We say that a vector field $\mathbf{f}$ (or the associated dynamical system)
is \emph{scale invariant} if it satisfies 
\begin{equation}
\mathbf{f}(\mathbf{a}\tau ^{\mathbf{p}})=\tau ^{\mathbf{p}-\mathbf{1}}%
\mathbf{f}(\mathbf{a}).  \label{basic2}
\end{equation}%
More generally, a vector field is called \emph{weight-homogeneous} with 
\emph{weighted degree} $\mathbf{d}$ if there is a vector $\mathbf{p}\in 
\mathbb{C}^{n}\setminus \{\mathbf{0}\}$, called \emph{the weight}, and a
vector $\mathbf{d}=(d_{1},\cdots ,d_{n})$ such that 
\begin{equation}
\mathbf{f}(\mathbf{a}\tau^{\mathbf{p}})=\tau ^{\mathbf{d}}\mathbf{f}(%
\mathbf{a}).
\end{equation}%
We denote the degree by $\textrm{deg}(\mathbf{f},\mathbf{p})=\mathbf{d}$. When 
$\mathbf{p}=\mathbf{1}$ the vector field is called \emph{homogeneous} of
degree $\mathbf{d}$. Note that any scale invariant vector field is a
weight-homogeneous field $\mathbf{f}=(f_{1},\cdots ,f_{n})$ such that each
component $f_{i}$ has degree $p_{i}-1$, from which it follows that $\partial
f_{i}/\partial x_{j}$ has weight $p_{i}-1-p_{j}$. If $\mathbf{f}$ has degree 
$\mathbf{d}$, then $\partial \mathbf{f}/\partial x_{j}$ has degree $\mathbf{d%
}-p_{j}\mathbf{e}_{j}$, with $\mathbf{e}_{j}$ the $j$-th unit vector. There
are weight-homogeneous vector fields that are not scale invariant; for
instance, all linear dynamical systems with constant coefficients which are
homogeneous vector fields.

Using Eqs. (\ref{basic1}), (\ref{basic2}) we see that given any nonzero
vector $\mathbf{p}$, a scale-invariant vector field $\mathbf{f}$ admits a
scale-invariant integral curve provided that the inhomogeneous linear system 
\begin{equation}
\mathbf{a}\mathbf{p}=\mathbf{f}(\mathbf{a}),
\end{equation}%
has nontrivial solutions for $\mathbf{a}$. Note that there may be nontrivial
solutions for $\mathbf{a}$ with some components zero. When at least one of
the components of $\mathbf{a}$ is nonzero, the corresponding solution of the
form (\ref{scale function}) is a particular solution of the dynamical
system. Therefore, in this case, we know the exact asymptotic behaviour of
the vector field $\mathbf{f}$ in the neighborhood of a finite-time
singularity, given by the solution (\ref{scale function}) with suitable $%
\mathbf{a}$'s and $\mathbf{p}$'s.

\section{Vector-field decompositions}

Unfortunately, most vector fields are neither scale invariant nor
weight-homogeneous. However, since any \emph{analytic} vector field $\mathbf{%
f}(\mathbf{x})$ can be expanded in a power series in some domain $\mathcal{D}%
\subset \mathcal{M}^{n}$, by taking any $\mathbf{x}_{0}\in \mathcal{D}$ for
any $\mathbf{x}\in \mathcal{D}$ distinct from $\mathbf{x}_{0},$ any such
vector field can be decomposed into weight-homogeneous \emph{components} by
taking for instance the first $k+1$ terms in its Taylor expansion around $%
\mathbf{x}_{0}$. A simpler example of a vector field admitting a
weight-homogeneous splitting is to take $\mathbf{f}$ to be any polynomial
vector field; then, a possible decomposition is to split it such that each
component is a suitable combination of monomials.

We say that the nonlinear vector field $\mathbf{f}$ on $\mathcal{M}^{n}$
admits a \emph{weight-homogeneous decomposition} with respect to a given
vector $\mathbf{p}$ if it splits as a combination of the form 
\begin{equation}
\mathbf{f}=\mathbf{f}^{\,(0)}+\mathbf{f}^{\,(1)}+\cdots +\mathbf{f}^{\,(k)},
\label{dec1}
\end{equation}%
where the \emph{components} $\mathbf{f}^{\,(j)},\,j=0,\cdots ,k,$ are
weight-homogeneous vector fields, namely, 
\begin{equation}
\mathbf{f}^{\,(j)}(\mathbf{a}\tau ^{\mathbf{p}})=\tau ^{\mathbf{p}+\mathbf{1}%
(q^{(j)}-1)}\mathbf{f}^{\,(j)}(\mathbf{a}),\quad j=0,\cdots ,k,  \label{dec2}
\end{equation}%
for some non-negative numbers $q^{(j)}$ and all $\mathbf{a}$ in some domain $%
\mathcal{E}$ of $\mathbb{R}^{n}$. In terms of individual components,
condition (\ref{dec2}) reads 
\begin{equation}
f_{\,i}^{\,(j)}(\mathbf{a}\tau ^{\mathbf{p}})=\tau
^{p_{i}+q^{(j)}-1}f_{\,i}^{\,(j)}(\mathbf{a}),\quad i=1,\cdots
,n,\,\,j=0,\cdots ,k.  \label{dec3}
\end{equation}%
In a slightly vague but suggestive manner we can say that a
weight-homogeneous decomposition splits the original vector field in parts
starting by collecting together the most nonlinear part and then proceeding
down to the `weakest' component such that each term in the splitting is
`less nonlinear' than the previous one in a precise sense.

There are two important features of such a vector-field decomposition (\ref%
{dec1}), (\ref{dec2}). Firstly, it is not unique. In general, for the given
vector field $\mathbf{f}$, many different vectors $\mathbf{p}$ can be found
which each lead to different weight-homogeneous decompositions of the vector
field in the same domain $\mathcal{D}$. Secondly, since the \emph{%
subdominant exponents} $q^{(j)},\,j=1,\cdots ,k$ can be ordered, 
\begin{equation}
0=q^{(0)}<q^{(j_{1})}<q^{(j_{2})},\quad \textrm{when}\quad j_{1}<j_{2},
\end{equation}%
the degrees of the component vector fields in the decomposition (\ref{dec2})
are also ordered 
and (only) the first vector field $\mathbf{f}^{\,(0)}$ appearing in the
decomposition is scale invariant.

Therefore, a weight-homogeneous decomposition of a vector field with respect
to a vector $\mathbf{p}$ is a splitting into $k+1$ weight-homogeneous
components each with degree $\mathbf{p}+\mathbf{1}(q^{(j)}-1),\,j=0,\cdots
,k,$ such that the lowest-order vector field in the decomposition is scale
invariant. The lowest-order vector field, $\mathbf{f}^{\,(0)}$, in a
splitting is sometimes called \emph{the dominant part} of $\mathbf{f}$ and
includes the most nonlinear terms in $\mathbf{f}$, whereas the remaining sum
of parts, $\mathbf{f}^{\textrm{sub}}\equiv \sum_{j=1}^{k}\mathbf{f}^{\,(j)}$,
is called \emph{the subdominant part}.

Given a vector field $\mathbf{f,}$ it is very important to have the complete
list of all possible weight-homogeneous decompositions it admits; in other
words, to know all the possible dominant and subdominant ways it can be
split. The asymptotic method we employ to trace the behaviour of vector
fields and their integral curves in a neighborhood of a movable finite-time
singularity (of a particular or  general solution), begins by finding all
weight-homogeneous decompositions of the vector field valid in that
neighborhood.

\section{The dominant balance}

Suppose that there exists a decomposition 
\begin{equation}
\mathbf{f}=\mathbf{f}^{\,(0)}+\mathbf{f}^{\,(1)}+\cdots +\mathbf{f}^{\,(k)},
\label{dec1'}
\end{equation}%
into ($k+1$) weight-homogeneous components, such that the dominant part $%
\mathbf{f}^{\,(0)}$ is scale invariant, and each subdominant component $%
\mathbf{f}^{\,(j)},\,j=1,\cdots ,k$ is weight-homogeneous. The scale
invariant solution 
\begin{equation}
\mathbf{x}^{(0)}(\tau )=\mathbf{a}\tau ^{\mathbf{p}},\,\,\mathbf{a}\neq 
\mathbf{0},\,\mathbf{p}\in \mathbb{Q}^{n},  \label{dec2'}
\end{equation}%
is a solution\footnote{%
due to a theorem of Goriely and Hyde \cite{gh1}, \cite{gh2}, we can restrict
attention to real vectors $\mathbf{a}$ only, as complex ones do not describe
the behaviour near finite time singularities in the sense of Eq. (\ref{sing2}%
).} of the dominant part $\dot{\mathbf{x}}=\mathbf{f}^{\,(0)}$ of the vector
field provided that $\mathbf{a}\mathbf{p}=\mathbf{f}^{\,(0)}(\mathbf{a})$.
Thus, some of the components of the vector $\mathbf{a}$ may vanish. We call
the components of the vector $\mathbf{p}=(p_{1},\cdots ,p_{n})$ the \emph{%
dominant exponents.} In this case, we sometimes say that (\ref{dec2'}) is an 
\emph{asymptotic solution} of the original system $\dot{\mathbf{x}}=\mathbf{f%
}(\mathbf{x})$.

On the other hand, the $k$ subdominant components $\mathbf{f}^{\,(j)}$
satisfy 
\begin{equation}
f_{\,i}^{\,(j)}(\mathbf{a}\tau ^{\mathbf{p}})=\tau
^{p_{i}+q^{(j)}-1}f_{\,i}^{\,(j)}(\mathbf{a}),\quad i=1,\cdots
,n,\,\,j=1,\cdots ,k,  \label{dec3'}
\end{equation}%
with the subdominant exponents $q^{(j)},\,j=1,\cdots ,k,$ which are ordered
and strictly positive. Dividing both sides by $\tau ^{\mathbf{p}-\mathbf{1}}$
and taking the limit as $\tau \rightarrow 0$, the true meaning of the
subdominant exponents is revealed, and we have 
\be
\lim_{\t\rightarrow
0}\frac{\mathbf{f}^{\textrm{sub}}(\bfA\t^{\bfp})}{\t^{\mathbf{p}-\mathbf{1}}}=0,
\ee
which proves that the subdominant part of the vector field,
$\mathbf{f}^{\textrm{sub}}$, is less dominant than the dominant
part, $\mathbf{f}^{(0)}$, which of course asymptotes as $\t^{\bfp-\mathbf{1}}$.

We say that the pair $(\mathbf{a},\mathbf{p})$ is a \emph{(dominant) balance}
for the vector field $\mathbf{f}$ if the latter admits a decomposition
satisfying Eqs. (\ref{dec1'})-(\ref{dec3'}). There may be several different
balances for any particular decomposition of $\mathbf{f}$ and a way to
classify them is by means of their order. The \emph{order} of a balance $(%
\mathbf{a},\mathbf{p})$ is the number of the nonzero components of the
vector $\mathbf{a}$. For a vector field on $\mathcal{M}^{n}$, the highest
order of any possible balance is $n$ and in this case the scale-invariant
solution (\ref{dec2'}) corresponds to a possible dominant behaviour of a 
\emph{general} solution of the original system $\dot{\mathbf{x}}=\mathbf{f}$
near the singularity. On the other hand, balances of a lower order than $n$
describe possible asymptotics of \emph{particular} solutions.

There is an elegant convex-geometric explanation of the dominant balances of
a vector field $\mathbf{f}$. This requires us first to express the vector
field in the so-called quasi-monomial form. Then, a dominant balance of
order $d$ corresponds precisely to a $d$-dimensional face of the
Newton-Puiseux-Bruno polyhedron associated to $\mathbf{f}$ (cf. \cite{br1}, 
\cite{br2}).

\section{The K-matrix}

We can now study an important square matrix, the so-called \emph{%
(K-)ovalevskaya matrix}, associated with a given vector field $\mathbf{f}$.
Consider the dominant part, $\mathbf{f}^{(0)}$, of $\mathbf{f}$ which admits
an exact solution of the form (\ref{dec2'}), described by the dominant
balance $(\mathbf{a},\mathbf{p})$ (of any nonzero order). The \emph{K-matrix
of the vector field $\mathbf{f}$ at the balance} $(\mathbf{a},\mathbf{p})$
is the square matrix 
\begin{equation}
\mathcal{K}=D{\mathbf{f}}^{(0)}(\mathbf{a})-\textrm{diag}\,\mathbf{p}.
\end{equation}%
The \emph{(K-)ovalevskaya exponents} associated with the balance $(\mathbf{a}%
,\mathbf{p})$ are the $n$ eigenvalues $(\rho _{1},\cdots ,\rho _{n})$ of $%
\mathcal{K}$. When the order of the balance is $n$ ($a_{i}\neq 0$ \textrm{%
for all} $i=1,\cdots ,n$), the \emph{K}-exponents are called the \emph{%
resonances} of $\mathcal{K}$.

Setting $\mathbf{w}=\mathbf{f}^{(0)}(\mathbf{a}\tau ^{\mathbf{p}}),$ and
differentiating with respect to $\tau ,$ we have 
\begin{equation}
\dot{\mathbf{w}}=(\mathbf{a}\mathbf{p}\tau ^{\mathbf{p}-\mathbf{1}})^{\cdot
}=\textrm{diag}(\mathbf{p}-\mathbf{1})\mathbf{a}\mathbf{p}\,\tau ^{\mathbf{p}-%
\mathbf{2}},
\end{equation}%
while by the chain rule 
\begin{eqnarray}
\dot{\mathbf{w}} &=&D\mathbf{f}^{(0)}(\mathbf{a}\tau ^{\mathbf{p}})\mathbf{a}%
\mathbf{p}\,\tau ^{\mathbf{p}-\mathbf{1}}  \nonumber \\
&=&D\mathbf{f}^{(0)}(\mathbf{a})\mathbf{a}\mathbf{p}\,\tau ^{\mathbf{p}-%
\mathbf{2}},
\end{eqnarray}%
where the last equality is most easily understood if we expand the
derivative to obtain $\dot{w}_{i}=\tau ^{p_{i}-2}\,\mathbf{\nabla }%
f_{i}^{(0)}(\mathbf{a})\cdot \mathbf{a}\mathbf{p}$, for the $i$-th
component. Thus, from the last two equations, we find 
\begin{equation}
\mathcal{K}\mathbf{a}\mathbf{p}=-\mathbf{a}\mathbf{p};
\end{equation}%
that is, the K-matrix always has the vector $\mathbf{f}^{(0)}(\mathbf{a})$
as an eigenvector with eigenvalue equal to $\rho _{1}=-1$. We say that a
balance is \emph{hyperbolic} if the remaining $(n-1)$ K-exponents have
positive real parts.

Suppose now that we know all the eigenvectors $\mathbf{v}^{(i)}$ and
eigenvalues $(-1,\rho _{2},\cdots ,\rho _{n})$ of the K-matrix. By simply
inspecting the form of the two sides in the variational equation for the
dominant part of the vector field, namely, the equation 
\begin{equation}
\dot{\mathbf{w}}=D\mathbf{f}^{(0)}(\mathbf{a}\tau ^{\mathbf{p}})\mathbf{w},
\end{equation}%
we can write the set of fundamental solutions $\mathbf{w}^{(i)}$ of this
linear equation for $\mathbf{w}$ in the form 
\begin{equation}
\mathbf{w}^{(i)}=\mathbf{g}^{(i)}(\log \tau )\,\tau ^{\mathbf{p}+\rho _{i}},
\label{var}
\end{equation}%
where, depending on whether or not $\mathcal{K}$ is semi-simple, the $%
\mathbf{g}^{(i)}(\log \tau )$'s are the eigenvectors $\mathbf{w}^{(i)}$ or,
in general, polynomials in $\log \tau $. Hence, the solutions of the
variational equation will be appropriate sums of terms by the form (\ref{var}%
).

Therefore, any solution of the original system will be well approximated
(cf. \cite{hs74}, p.299, for the precise conditions) by a solution of the
form (this is a Taylor estimate): 
\begin{equation}
\mathbf{x}=\mathbf{x}^{(0)}+\mathbf{w},  \label{main1}
\end{equation}%
where 
\begin{equation}
\mathbf{x}^{(0)}(\tau )=\mathbf{a}\tau ^{\bfp},\quad \mathbf{w}%
=\sum_{i=1}^{k}\mathbf{h}^{(i)}\tau ^{\mathbf{p}+\rho _{i}},
\end{equation}%
where the $\mathbf{h}^{(i)}$'s are, in general, polynomials in $\log \tau $.
Furthermore, we arrive at the interesting conclusion that in any particular
or general solution, the arbitrary constants characterizing it will first
appear in those terms whose coefficients have indices equal to a K-exponent.
In a general solution, the arbitrary constants normally appear at different
places in an expansion, and consequently, a solution in which a K-exponent
(different from the $-1$ value which as we have shown always exists) is
either negative, or has nontrivial multiplicity, may or may not be a general
solution. We shall see in the next section how we can use Eq. (\ref{main1})
to obtain an important series representation of the solutions to the
original dynamical system near a finite-time singularity.

Apart from the principal use made here in unravelling the nature of
finite-time singularities, there are various important connections between
the K-exponents, first integrals, integrability properties of Hamiltonian
systems and complex algebraic geometry, cf. \cite{yo83}, \cite{avm1}.

\section{Formal expansions}

In the previous subsection, we derived the formal expansion (\ref{main1})
for the solution of the dynamical system $\dot{\mathbf{x}}=\mathbf{f}(%
\mathbf{x})$ near a finite-time singularity by including only the first
terms in a power-series expansion of $\mathbf{x}$. That solution corresponds
to a given dominant balance $\mathcal{F}=(\mathbf{a},\mathbf{p})$ and
depends on the K-exponents $(-1,\rho _{2},\cdots ,\rho _{n})$ associated
with the dominant part, $\mathbf{f}^{(0)}$, of the original vector field $%
\mathbf{f}$. Writing Eq. (\ref{main1}) in full, that is including terms of
all orders, can lead to very complicated expansions in general, and the
essential feature is the appearance of logarithmic terms. We note that there
are a number of general theorems guaranteeing the existence and convergence
of such expansions -- see, for instance, \cite{g01} for a review.

Let $\mathcal{E}^{+}$ be the set of all K-exponents with positive real
parts. For simplicity we assume that all K-exponents in $\mathcal{E}^{+}$
are rational, and define the number $1/s$ to be the least common multiple of
the denominators of the numbers in the set $\mathcal{H}=\{q^{(1)},\cdots
,q^{(m)}\}\cup \mathcal{E}^{+}$. In the case where any log terms are absent
(for instance, when K is semi-simple), we can write, by (\ref{main1}), the
full expansion of the general solution around the finite-time singularity in
the form of a \emph{Puiseux series}, \ 
\begin{equation}
\mathbf{x}=\tau ^{\mathbf{p}}\left( \mathbf{a}+\sum_{i=1}^{\infty }\mathbf{c}%
_{i}\tau ^{i/s}\right) ,  \label{main2}
\end{equation}%
where, as we know already from the previous section, each of the $n$
arbitrary constants in (\ref{main2}) will first appear in the term with
coefficient $\mathbf{c}_{k},\,k=\rho s$ and $\rho \in \mathcal{E}^{+}$.

Hence, finding the final form of the solution (general or particular,
depending on the number of arbitrary constants appearing in the series
expansion) can be now reduced to knowing the coefficients $\mathbf{c}_{i}$
in the expansion. These coefficients are computed by inserting the Puiseux
series (\ref{main2}) into the original system. This leads to a set of \emph{%
recursion relations}, a linear system for the coefficients $\mathbf{c}_{i}$.
For the $j$th-order coefficient we find 
\begin{equation}  \label{main3}
\mathcal{K}\mathbf{c}_{j}-\frac{j}{s}\mathbf{c}_{j}=\mathbf{P}_j (\mathbf{c}%
_{1},\cdots,\mathbf{c}_{j-1}),
\end{equation}
where the forms $\mathbf{P}_j$ are polynomial in its variable, read off from
the original equation.

There is an important consistency condition to be satisfied for the above
analysis to be valid. Multiplying both sides of (\ref{main3}) by $\mathbf{v}$%
, an eigenvector of the K-matrix, we see that when $j/s=\rho $, an
eigenvalue of the K-matrix, we must have the following \emph{compatibility
condition} ($\mathbf{v}^{\top }$ denotes the adjoint eigenvector of $\mathbf{%
v}$): 
\begin{equation}
\mathbf{v}^{\top }\cdot \mathbf{P}_{j}=0,\quad \textrm{for all}\,\,\rho \in 
\mathcal{E}^{+}.
\end{equation}%
Therefore, if the above compatibility condition is \emph{violated} at some
eigenvalue, then we conclude that no solution in the form of a Puiseux
series can exist and we have to search for more general solutions which may
contain logarithmic terms. Such a more general series will be of the form of
a \emph{$\psi $-series} (cf. \cite{bo79} for this terminology): a direct
generalization of the form (\ref{main2}) 
\be\label{main4}
\mathbf{x}=\t^\bfp\left(\mathbf{a}+\sum_{i=1}^{\infty}\sum_{j=1}^{\infty}
\mathbf{c}_{ij}\t^{i/s}(\t^\r\log\t)^{j/s}\right), \ee
where $\rho $ is the first K-exponent for which the compatibility condition
is not satisfied and $s$ as defined above. The procedure for the calculation
of the coefficients in this more general case is the same as before, leading
again to the form of the general solution in a suitable neighborhood of the
finite-time singularity as a $\psi $-series.

\section{Relation to unstable manifolds}

Suppose that (\ref{dec2'}) is an asymptotic solution of the vector field $%
\dot{\mathbf{x}}=\mathbf{f}(\mathbf{x})$. We set 
\begin{equation}
\mathbf{x}(\tau )=\tau ^{\textbf{p}}\mathbf{X}(s),\quad s=\log \tau ,
\end{equation}%
and imagine that an equation of the form 
\begin{equation}
\mathbf{X}(s)=\textrm{const.}=\mathbf{a},
\end{equation}%
regards the coefficient $\mathbf{a}$ of a given balance $(\mathbf{a},\mathbf{%
p})$ of the vector field $\mathbf{f}$ as an equilibrium point of the new
system given, in terms of the new variables $\mathbf{X}(s)=(X_{1}(s),\cdots
,X_{n}(s))$, by 
\begin{eqnarray}
X_{i}^{\prime } &=&F_{i}(X_{1},\cdots ,X_{n}),\quad i=1,\cdots ,n  \nonumber
\label{comp} \\
X_{n+1}^{\prime } &=&qX_{n+1},\quad X_{n+1}=e^{qs},
\end{eqnarray}%
where $(^{^{\prime }})=d/ds$ and $q$ is the least common multiple of the
denominators of the subdominant exponents in the original system. We call
the dynamical system (\ref{comp}) the \emph{companion system } of the
original vector field $\dot{\mathbf{x}}=\mathbf{f}(\mathbf{x})$. Thus, the
transformation (\ref{comp}) to the companion system associates a different
system of this sort to each one of the balances $(\mathbf{a},\mathbf{p})$ of
the original system provided that any given pair of balances has different
dominant exponents $\mathbf{p})$.

Consider now the \emph{linearized system} 
\begin{equation}
\mathbf{X}^{\prime }=D\mathbf{F}(\mathbf{a})\mathbf{X},
\end{equation}%
which is the variational equation of the companion system around the
equilibrium point $\mathbf{X}=(\mathbf{a},0)$. This is a
constant-coefficient, linear system. From the fundamental theorem of such
systems, it follows that the general solution passing through the initial
condition $\mathbf{X}(0)=\mathbf{X}_{0}$ is given by 
\begin{equation}
\mathbf{X}(s)=e^{sD\mathbf{F}(\mathbf{a})}\mathbf{X}_{0}.
\end{equation}%
Now we know (cf. \cite{ta}) that for any $n\times n$ matrix $\mathbf{A}$: 
\begin{equation}
e^{s\mathbf{A}}\mathbf{v}=\sum e^{\lambda _{j}s}v_{j}(s),
\end{equation}%
where $\lambda _{j}$ are the eigenvalues of $\mathbf{A}$ and $v_{j}$ are $%
\mathbb{C}^{n}$-valued polynomials, the latter being constant if and only if 
$A$ is semi-simple (diagonalizable). So 
\begin{equation}
\mathbf{X}=\sum v_{j}(\log \tau )\tau ^{\lambda _{j}},
\end{equation}%
in terms of the $\tau $-time. Hence, using this form of the local solution
to the companion system, we can express the local solution to the original
system around its finite-time singularity in the form 
\begin{equation}
\mathbf{x}(\tau )=\sum v_{j}(\log \tau )\tau ^{\mathbf{p}+\lambda _{j}},
\end{equation}%
which is precisely the Taylor estimate leading eventually to the $psi$%
-series representation we arrived at in previous Sections. We therefore
reach the interesting conclusion that around the equilibrium point $(\mathbf{%
0},q)$, origin of the companion system (\ref{comp}), the eigenvalues of that
system are simply $(\mathbf{p},q)$; that is, the dominant exponents of the
asymptotic solution of the original system together with the number $q$
characterizing the subdominant part of the vector field. Moreover, around
the equilibrium point $(\mathbf{a},q)$ of the companion system (\ref{comp}),
the eigenvalues $\lambda _{1},\cdots ,\lambda _{j},q$ of the companion
system are $(\rho _{1},\cdots ,\rho _{n},q)$; that is, they are precisely
the K-exponents of the original system together with the subdominant number $%
q$.

We have an  interpretation of the results of the previous Sections
concerning the local behaviour of the original system around its finite-time
singularities from a dynamical systems perspective. Using the companion
transformation, a local analysis of the companion system $\mathbf{X}^{\prime
}=\mathbf{F}(\mathbf{X})$ \emph{around its equilibrium points} (with $%
s\rightarrow -\infty $ necessarily) will provide the local analysis of the
solutions of all possible balances of the original system \emph{around its
singularities} ($\tau =e^{qs}\rightarrow 0$ since $q>0$ always, for an
acceptable decomposition).

Note that, since we need $s\rightarrow -\infty $, we are only interested in
the unstable manifold (eigenvalues with positive real parts) of the
equilibrium points of the companion system. Therefore, the negative
K-exponents (corresponding to the stable manifold of the companion system)
are not connected to the behaviour of solutions of the original system at
the finite-time singularity, but are associated with its behaviour as $\tau
\rightarrow \infty $.

For a discussion of the companion transformation in connection with
integrability and complex dynamics, see \cite{g01} and Refs. therein.

\section{The method of asymptotic splittings}

\label{method} The results in the previous subsections suggest a general
procedure to uncover the nature of singularities by constructing series
expansion representations of particular or general solutions of dynamical
systems in suitable neighborhoods of their finite-time singularities. This
method consists of building splittings of vector fields that are valid
asymptotically and trace the dominant behaviour of the vector field near the
singularity. A resulting series expansion connected to a particular dominant
balance helps to decide whether or not the arrived solution is a general one
and to spot the exact positions of the arbitrary constants as well as their
role in deciding about the nature of the time singularity.

The method we suggest below is analogous to the so-called \textsc{ARS}
procedure connected with the Painlev\'{e} and integrability properties of
dynamical systems \cite{ars}, but here the whole approach and viewpoint are
completely different. We are not concerned with notions of integrability but
solely with the problem of the nature of finite-time singularities. Also,
our systems are real-valued with a real time variable.

To apply this \emph{method of asymptotic splittings} to a particular
dynamical system in an effort to discover the nature of its time
singularities, we must follow this recipe:

\begin{enumerate}
\item Write the system of equations in the form of a dynamical system $\dot{%
\mathbf{x}}=\mathbf{f}(\mathbf{x})$ with $\mathbf{x}=(x_1,\cdots,x_n)$, and
identify the vector field $\mathbf{f}(\mathbf{x})=(f_1(\mathbf{x}%
),\cdots,f_n(\mathbf{x}))$.

\item Find all the different weight-homogeneous decompositions of the
system; that is, the splittings of the form 
\[
\mathbf{f}=\mathbf{f}^{\,(0)}+\mathbf{f}^{\,(1)}+\cdots +\mathbf{f}^{\,(k)},
\]%
and choose one of these splittings to start the procedure.

\item Substitute the scale-invariant solution 
\[
\mathbf{x}^{(0)}(\tau )=\mathbf{a}\tau ^{\mathbf{p}},
\]%
into the equation $\dot{\mathbf{x}}=\mathbf{f}^{(0)}$. Study the resulting
algebraic systems, and find all dominant balances $(\mathbf{a},\mathbf{p})$
together with their orders.

\item Identify the non-dominant exponents, that is the positive numbers
$q^{(j)},\, j=1,\cdots,k$, such that
 \[
\t^{q^{(j)}}\sim\frac{\textbf{f}^{\,\,\textrm{sub},\,{(j)}}(\t^{\bfp})}
{\t^{\mathbf{p}-\mathbf{1}}}\rightarrow 0.
\]

\item Construct the K-matrix $\mathcal{K}$: 
\[
\mathbf{f}^{(0)}\rightarrow D\mathbf{f}^{(0)}\rightarrow D\mathbf{f}^{(0)}(%
\mathbf{a})\rightarrow D\mathbf{f}^{(0)}(\mathbf{a})-\textrm{diag}\,\mathbf{p}
. 
\]

\item Compute the spectrum of $\mathcal{K}$, 
\[
\textrm{spec}(\mathcal{K})=(-1,\rho _{2},\cdots ,\rho _{n}). 
\]%
Is $\mathcal{K}$ semi-simple? Are the balances hyperbolic?

\item Find the eigenvectors $\mathbf{v}^{(i)}$ of $\mathcal{K}$.

\item Identify $s$ as the multiplicative inverse of the least common
multiple of all the subdominant exponents and positive K-exponents.

\item \label{puiseux} Substitute the Puiseux series 
\[
x_{i}=\sum_{j=0}^{\infty }c_{ji}\,\tau ^{p_{i}+\frac{j}{s}}
\]%
into the original system.

\item Identify the polynomials $\mathbf{P}_{j}$ and solve for the final
recursion relations which give the unknown coefficients $\mathbf{c}_{j}$.

\item Check the compatibility conditions at the K-exponents, 
\[
\mathbf{v}^{\top}_\r\cdot\mathbf{P}_\r=0,\quad \textrm{for each eigenvalue}%
\,\,\rho. 
\]

\item If the Puiseux series is valid, then the method is concluded for this
particular splitting. Otherwise, if compatibility conditions are violated at
the eigenvalue $\rho ^{\ast }$, restart from step \ref{puiseux} by
substituting the logarithmic series (\ref{main4}).

\item Get coefficient at order $\rho ^{\ast }$. Write down the final
expansion with terms up to order $\rho ^{\ast }$.

\item Verify that compatibility at $\rho^*$ is now satisfied.

\item Repeat whole procedure for each of the other possible decompositions.
\end{enumerate}

Note that although the whole spectrum of possible behaviours of the system
near a time singularity is concluded once we find valid series expansions
corresponding to each balance in each particular decomposition, an
additional analysis of the phase space of the companion systems
corresponding to each one of the balances of the original system may lead to
valuable insights as to the geometric structure of the phase space -- \emph{%
how the orbits behave} -- in the neighborhood of the finite-time singularity.

Following the above steps even up to that of calculating a dominant balance
in one particular decomposition, can be very useful since it gives you   one
particular possible asymptotic behaviour of the system near the time
singularity. In this respect, the whole method expounded here is truly
generic since it helps to decide the generality of any behaviour found in an
exact solution -- that is, how many arbitrary constants there are in the
final solution that shares that behaviour (particular or general solution).
It is rare that a Puiseux series is inadequate to describe the dynamics
(semi-simplicity of K), but in such uncommon cases one must resort to the
more complex logarithmic solutions.

\section{A worked example}

As a concrete example of the above analysis, consider the homogeneous and
isotropic FRW cosmological equations in general relativity for a perfect
fluid of pressure $p(t)$, density $\rho (t),$ and equation of state $p=w\rho 
$, where $w$ is a constant. They read 
\begin{eqnarray}
\frac{\ddot{a}}{a} &=&-\frac{\rho }{6}(1+3w),  \label{2a} \\
\dot{\rho}+3H\rho (1+w) &=&0, \\
3H^{2} &=&\rho -\frac{k}{a^{2}},  \label{3a}
\end{eqnarray}%
where $a(t)$ is the scale factor, $H(t)$ the Hubble expansion rate and $k$
the constant spatial curvature. Setting $a=x,\dot{x}=y,\rho =z$, this system
reads 
\begin{eqnarray}
\dot{x} &=&y,  \label{4} \\
\dot{y} &=&axz,  \label{5} \\
\dot{z} &=&byz/x,  \label{6}
\end{eqnarray}%
with 
\begin{equation}
a=-(1+3w)/6,\quad b=-3(1+w),  \label{7}
\end{equation}%
and is subject to the integral constraint (the Friedmann equation) 
\begin{equation}
3\frac{y^{2}}{x^{2}}+\frac{k}{x^{2}}=z.  \label{8}
\end{equation}%
This system is weight-homogeneous. The unique balance is determined by 
\begin{equation}
\mathbf{a}=(\alpha ,\beta ,\gamma )=\left( \alpha ,p\alpha ,\frac{2(2+b)}{%
ab^{2}}\right) ,\quad \mathbf{p}=(p,q,r)=\left( -\frac{2}{b},-\frac{2}{b}%
-1,-2\right) ,
\end{equation}%
and the parameters $a,b$ depend on $w$ via Eq. (\ref{7}). Since the field is
weight-homogeneous (i.e., $\mathbf{f}=\mathbf{f}^{(0)}$), this balance
corresponds to an exact, scale-invariant solution of the original system.

Note that there is one arbitrary coefficient, and so we expect that one of
the K-exponents will be zero. For the vector field $\mathbf{f}=(y,axz,byz/x)$
the associated K-matrix, $K=D\mathbf{f}(\mathbf{a})-\textrm{diag}\mathbf{p},$
has characteristic equation 
\begin{equation}
r^{3}+(2p-1)r^{2}+2(p-1)r=0,
\end{equation}%
hence the K-exponents are 
\begin{equation}
r=-1,\,\,0,\,\,2(1-p).  \label{13}
\end{equation}%
We can use the integral constraint (\ref{8}) to get a value for the
coefficient $\alpha $. Balancing the terms in (\ref{8}) leads to 
\begin{equation}
\alpha =\pm \sqrt{\frac{3\beta ^{2}ab^{2}}{2(2+b)}}.
\end{equation}%
In the case when all components of the vector $\mathbf{a}$ are real, the
solution $X=(x,y,z)\equiv (a,\dot{a},\rho )$ of the dynamical system
experiences a finite-time singularity. When $1-p>0$, that is when either $%
w<-1$ or $w>-1/3$, solutions are general; while, in the range $-1<w<-1/3,$
we have only behaviours corresponding to particular solutions of the system.
Note that when $w=0$, we have $b=-3$ and we find a behaviour similar to that
of dust, but when $w=1/3$, $b=-4$ the behaviour is that of the standard
radiation models. Further, calculating the recursion relations to compute
the coefficients of the series expansion term by term, we find the following
asymptotic solution for radiation, 
\begin{eqnarray*}
x &=&\alpha \tau ^{1/2}-\frac{\alpha }{3}\tau ^{3/2}-\frac{\alpha }{18}%
c_{31}\tau ^{5/2}+\cdots  \\
z &=&\frac{3}{4}\tau ^{-2}+c_{31}\tau ^{-1}+c_{31}^{2}+\frac{8}{9}%
c_{31}^{3}\tau +\cdots ,
\end{eqnarray*}%
while the dust-dominated expansion is found to be 
\begin{eqnarray*}
x &=&\alpha \tau ^{2/3}-\frac{\alpha }{4}\tau ^{4/3}+\cdots  \\
z &=&\frac{4}{3}\tau ^{-2}+c_{32}\tau ^{-4/3}+\cdots 
\end{eqnarray*}

The method that we have described in detail here provides a toolkit for the
investigation of the general form of a range of finite-time singularities in
general relativistic cosmologies. In particular, the 'sudden' singularities
introduced by one of us \cite{jdb}, in which $a,\dot{a},$ and $\rho $ remain
finite but $\ddot{a}$ and $p\rightarrow \infty $ at finite time in
situations where no functional relationship is assumed between $p$ and $\rho 
$, have been widely studied \cite{sud}. Such situations, allow sudden
singularities to develop at finite time without violating the strong-energy
condition of general relativity and are require less severe conditions than
future 'big rip' singularities \cite{rip} with $w<-1$. Elsewhere, we will
report on the results of applying these methods to determine the general
behaviour on approach to these singularities in general relativity and in
higher-order gravity theories.

\ack We thank Yvonne Choquet-Bruhat, Alain Goriely and Antonis Tsokaros for
useful discussions. The work of S.C. was partly supported by the Ministry of
Education and Religious Affairs (25\%) and by E.U.(75\%) under the Grant
"Pythagoras".

\end{document}